\documentclass{eas}
\usepackage{graphicx}
\begin{document}


\title{SOLAR LIMB DARKENING FUNCTION AND SOLAR DIAMETER WITH ECLIPSE OBSERVATIONS} 

\runningtitle{Limb Darkening Function with Eclipse Observations}

\author{A. Raponi}
\address{Sapienza University of Rome, Italy. email: andr.raponi@gmail.com}
\author{C. Sigismondi}
\address{Sapienza University of Rome and ICRA, University of Nice-Sophia Antipolis, IRSOL and GPA-Observatorio Nacional Rio de Janeiro.}

\author{K. Guhl}\author{R. Nugent}\author{A. Tegtmeier}
\address{IOTA, International Occultation Timing Association}

\begin{abstract}

We introduce a new method to perform high resolution astrometry of the solar diameter from the ground, through the observations of eclipses. A discussion of the solar diameter and its variations is linked to the Limb Darkening Function (LDF) using the luminosity evolution of a Baily's Bead and the profile of the lunar limb available from satellite data. The inflexion point of the LDF is defined as the solar limb.
The method proposed is applied for the videos of the eclipse in January, 15, 2010 recorded by Richard Nugent in Uganda and Andreas Tegtmeier in India. An upper limit for the inflexion point position has been set for that eclipse.
 
\end{abstract}

\maketitle


\section{The method of eclipses}

With the eclipse observation we are able to bypass some problems that affect the measurement of solar diameter. The atmospheric and instrumental effects that distort the shape of the limb [\cite{Djafer}] are overcome by the fact that the scattering of the Sun's light is greatly reduced by the occultation of the Moon, therefore there are much less photons from the photosphere to be poured, by the PSF effect, in the outer region. 

The method exploits the observation of the beads of light that appear or disappear from the bottom of a lunar valley when the solar limb is almost tangent to the lunar limb. 

The shape of the light curve of the bead is determined by the shape of the LDF (not affected by seeing) and the shape of the lunar valley that generates the bead. 
Calling w(x) the width of the lunar valley (i.e. the length of the solar edge visible from the valley in function of the height x from the bottom of the valley), and B(x) the surface brightness profile (i.e. the LDF), one could see the light curve L(y) as a convolution of B(x) and w(x), being $\mid y\mid$ the distance between the botton of the lunar valley and the standard solar edge, setting to 0 the position of the standard edge.

\noindent $L(y)=\intop B(x)\, w(y-x)\, dx$

\noindent The discrete convolution is:

\noindent $L(m)=\sum B(n)\, w(m-n) h$

\noindent where n, m are the index of the discrete layers corresponding to x, y coordinate and h the layers thickness.

The profile of the LDF is discretized in order to calculate the deconvolution and thus to obtain the solar layers B(n) (see Fig. 1).

\begin{figure}
\centerline{\includegraphics[width=0.6\textwidth,clip=]{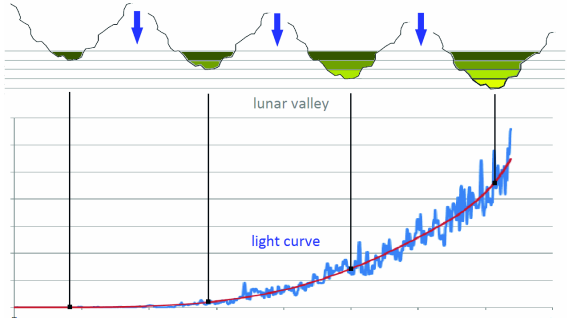}}
\caption{Every step in the geometry of the solar-lunar layers (up) corresponds to a given instant in the light curve (down). The value of the light curve is the contribute of all the layers.}
\label{Fig. 1}
\end{figure}

\section{An application of the method}

We studied the videos of the annular eclipse in January 15, 2010 realized by Richard Nugent in Uganda and Andreas Tegtmeier in India with CCD camera Watec and Matsukov telescope.
Two beads located at Axis Angle\footnote[1]{the angle around the limb of the Moon, measured Eastward from the Moon's North pole} (AA) $=171^\circ, 177^\circ$ are analyzed for both the videos. 
The lunar valley analysis is performed with the Occult 4 software \footnote[2]{$www.lunar-occultations.com/iota/occult4.htm$} that exploits the new lunar profile obtained by the laser altimeter (LALT) onboard the Japanese lunar explorer Kaguya \footnote[3]{$http://wms.selene.jaxa.jp/selene\_viewer/index\_e.html$}. The choice of the thickness of the layers has to be optimal: large enough to reduce $B_n$ uncertainties, but small enough to have a good resolution of the LDF. 
The resulting points show the profiles obtained in two different positions. The inflection point is clearly between the two profiles (see Fig. 2).

\begin{figure}
\centerline{\includegraphics[width=0.7\textwidth,clip=]{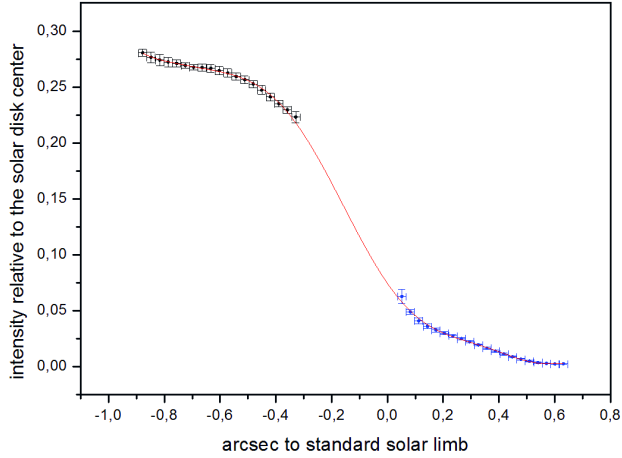}} 
\caption{The luminosity profiles obtained for the bead at AA = 177$^\circ$ are plotted and put together. The inner and brighter part is obtained from Tegtmeier's video; the outer and weaker part is obtained from Nugent's video. The luminosity profile is normalized to the center of the solar disk according to Rogerson [\cite{Rogerson}] for the inner part, and in arbitrary way for the outer part. The zero of the abscissa is the position of the standard solar limb with a radius of 959.63 arcsec at 1 AU. The error bars on y axis are the 90\% confidence level. The error bars on x axis are the thickness (h) of the lunar layers. The solid line is an interpolation between the profiles and gives a possible scenario on the position of the inflection point.}
\label{Fig. 2}
\end{figure}

\section{Conclusions}

This study takes into account the potentiality of the observation of eclipses in defining the luminosity profile of the edge of the Sun. 

The method proposed considers the bead as a light curve forged by the LDF and the profile of the lunar valley. A first application on two beads of the annular eclipse on 15 January 2010, is described in this study. We obtain a detailed profile, demonstrating the functionality of the method. Although it was impossible to observe the inflection point, its position is defined within a narrow range. The solar radius is thus defined within this range (from -0.19 to +0.05 with respect to the standard radius), resulting compatible with the standard value.

%

\end{document}